\documentclass{osa-article}

\journal{oe}

\begin{document}

\title{Continuous spatial field confocal thermometry using lanthanide doped tellurite glass}

\author{D. Stavrevski,\authormark{1,*} E. Schartner,\authormark{2,3} I. S. Maksymov,\authormark{4} R. A. McLaughlin,\authormark{3,5} Heike Ebendorff-Heidepriem,\authormark{2,3} and A. D. Greentree\authormark{1}}

\address{\authormark{1}Australian Research Council Centre of Excellence for Nanoscale BioPhotonics, School of Science, RMIT University, Melbourne, VIC 3001, Australia\\
\authormark{2}Australian Research Council Centre of Excellence for Nanoscale BioPhotonics, School of Physical Sciences, The University of Adelaide, Adelaide, SA 5005, Australia\\
\authormark{3}Institute for Photonics and Advanced Sensing, The University of Adelaide, Adelaide SA 5005, Australia
\authormark{4}Centre for Micro-Photonics, Swinburne University of Technology, Hawthorn, VIC 3122, Australia\\
\authormark{5}Australian Research Council Centre of Excellence for Nanoscale BioPhotonics, Adelaide Medical School, The University of Adelaide, Adelaide, SA 5005, Australia }
\email{\authormark{*}daniel.stavrevski@rmit.edu.au} 




\begin{abstract}
Quantifying temperature variations at the micron scale can provide new opportunities in optical sensing. In this paper, we present a novel approach using the temperature-dependent variations in fluorescence of rare-earth doped tellurite glass to provide a micron-scale image of temperature variations over a 200 micrometre field of view. We demonstrate the system by monitoring the evaporation of a water droplet and report a net temperature change of 7.04 K with a sensitivity of at least 0.12 K. These results establish the practicality of this confocal-based approach to provide high-resolution marker-free optical temperature sensing.
\end{abstract}

\section{Introduction}
Characterising small-scale temperature variations with respect to location and time is useful in both industrial and biological applications. When considering the breakdown of semiconductor devices, an initial 'hot spot' can cause thermal runaway\cite{popescu1970selfheating,shenai1989optimum} that can ruin en entire device. In monitoring of chemical reactions, self-sustaining reactions can be initiated from the combination of two molecules\cite{hordijk2004detecting}. In biological systems, temperature variation may be indicative of disease progression, metabolism\cite{chretien2018mitochondria} and inflammation\cite{toutouzas2011new,kokate2003intravascular}.

There are many different approaches that are used to provide accurate and highly-localized measures of temperature, depending on the characteristics of the system to be monitored. A standard solution is to implement fluorescent thermometers comprised of small organic compounds that are able to target specific regions, such as organelles in cells that play an important role in thermogenic processes\cite{arai2014molecular}. Mito thermo yellow and Mito-RTP\cite{arai2015mitochondria,homma2015ratiometric} are examples of fluorescent thermometers that are able to bind to the mitochondria for fast monitoring of thermal processes related to a cell's metabolism. Other fluorescent temperature sensors include quantum dots\cite{li2007single,benayas2015pbs,del2016infrared} and nitrogen vacancy centres in diamond\cite{kucsko2013nanometre} which have been developed to measure point-based temperature in various chemical and biological systems with sub-micron resolution. Thermal cameras based on near infrared imaging have been used to monitor heat generation of electro-chemical processes in fuel cells\cite{pomfret2010thermal}, with a spatial resolution of $\sim$ 0.1mm.

Previously Er$^{3+}$:Yb$^{3+}$ co doped tellurite glass (EYT) has been used for monitoring temperature in biological systems with high sensitivity ($\approx$ 10$^{-3}$/K) at temperature ranges of importance to biological processes\cite{schartner2014fibre,musolino2016portable,manzani2017portable}. These studies used a ratiometric approach to fluorescence temperature sensing in fibre based probes as a means to sense changes in temperature at discrete points in space with high sensitivity. The flexibilty of using fibres as temperature probes has also been demonstrated by monitoring  intracranial temperature of a rat\cite{musolino2016portable}, highlighting the biocompatibility of tellurite composites for biological applications. Here we present a non-standard approach to temperature sensing by using scanning confocal microscopy (SCM) to monitor the temperature sensitive upconversion fluorescence of a rare-earth doped glass. Our SCM approach  allows thermometry over a 200 $\mu$m field of view, with spatial resolution at the micron scale. We demonstrate this ability via a proof of concept experiment by monitoring the evaporative cooling of a water droplet on the substrate.

\section{Mechanism and design}
The temperature sensing mechanism of EYT involves a ratiometric change in the visible part of the fluorescence  spectrum as a result of variations in local temperature. This process can be triggered by using an excitation in the visible wavelength range (typically at either 476 and 514 nm)\cite{haouari2017optical} or in the near infrared region via two-photon upconversion. The high rare-earth solubility of this glass allows for the  incorporation of a high concentration\cite{vetrone2002980} of Yb$^{3+}$, which acts as a sensitiser in the EYT emission mechanism\cite{schartner2014fibre} resulting in a significant increase in fluorescence output.

\begin{figure}[t]
    \centering
    \includegraphics[width=\textwidth]{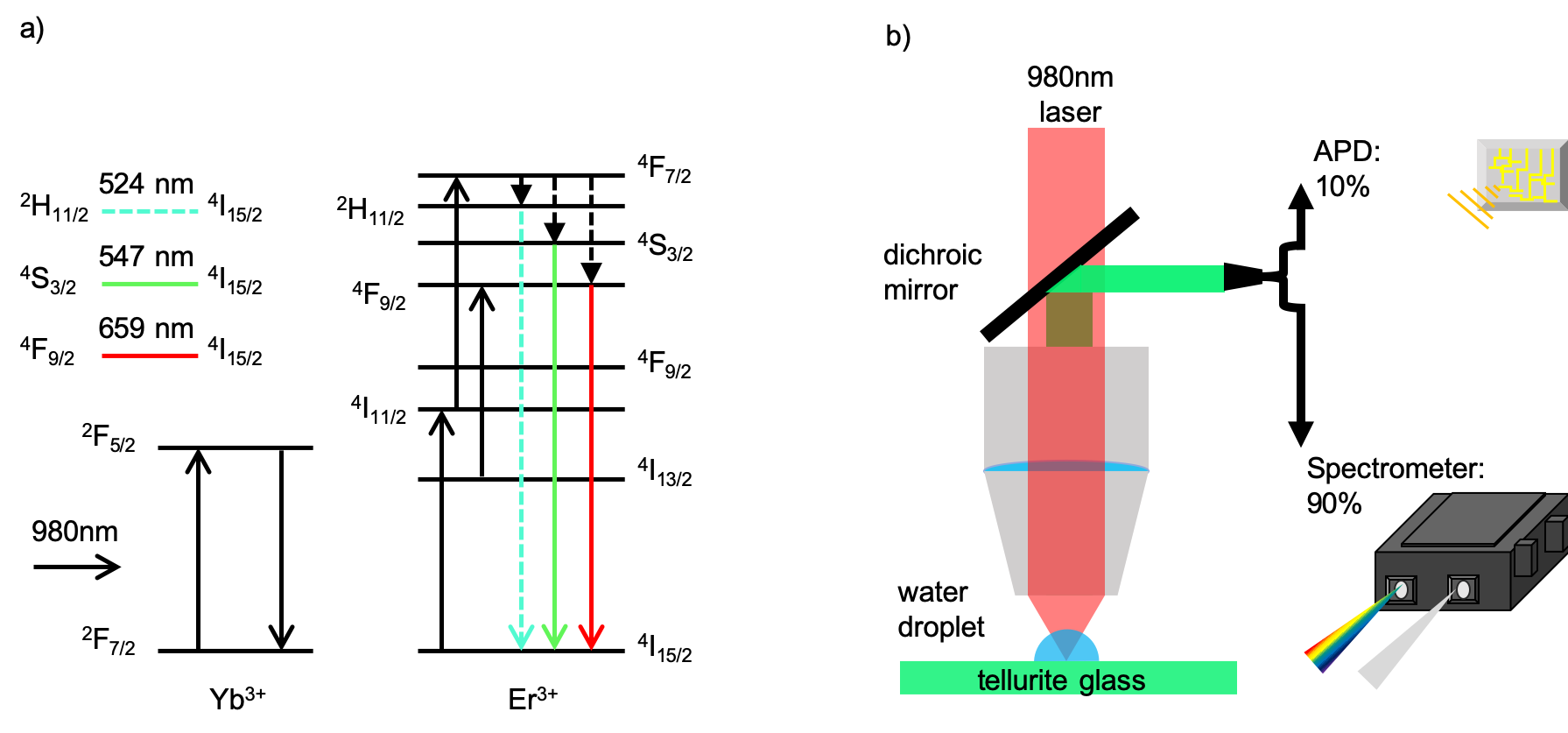}
    \caption{a) The energy level diagram for Er$^{3+}$ and Yb$^{3+}$, where Yb$^{3+}$ acts as a sensitizer in the upconversion process due to the overlap in excitation pathway with Er$^{3+}$ populating the $^{4}$I$_{11/2}$ state. A secondary excitation due to  simultaneous two-photon absorption results in photo emission in the visible wavelength range. The upconversion process governed by excitation at 980 nm yields two temperature dependent emission bands at 524 and 547 nm. b) A schematic of the confocal scanning microscopy setup used to measure temperature of a water droplet. The total output signal is split with 90 $\%$ directed towards a spectrometer for high sensitivity and 10 $\%$ to an APD for fast, real-time imaging.}
    \label{fig:1}
\end{figure}

The visible excitation pathway, used to measure features of proteins and tissues\cite{aubin1979autofluorescence,huang2004laser} results in auto-fluorescence effects which increases the overall background fluorescence, hence reducing sensitivity. Although we do not monitor biological tissue here, nevertheless we still use the infrared channel to understand the use of our sensor for future biological applications.  Accordingly, we investigate the temperature-dependent  fluorescence with 980 nm excitation. The 980 nm pump is absorbed by the Yb$^{3+}$ ions, which transfer the excitation to long-lived transitions in the proximal Er$^{3+}$, as shown in Fig. 1a.  These long-lived transitions facilitate the upconversion process. If we consider only the sequential two-photon absorption to the $^{4}$F$_{7/2}$ state, then phononic processes determine the decay to the $^{2}$H$_{11/2}$ and $^{4}$S$_{3/2}$ states.  Hence the relative contributions of emission from these states to the ground state, with centre wavelengths 524 nm and 547 nm respectively, serves as a probe for the temperature.

The EYT glass substrate was prepared using a previously reported method\cite{oermann2009index}, including doping of 1 $\times$ 10$^{20}$ cm$^{-3}$ erbium and 9 $\times$ 10$^{20}$ cm$^{-3}$ of ytterbium. The batch of raw materials was melted in a gold crucible at 830 $^{\circ}$C in open air for approximately 30 minutes and then cast onto a warm brass plate, which was allowed to cool to room temperature. This resulted in a thin glass piece of approximately 20$\times$15$\times$3mm of non-regular dimensions. The top and bottom faces were then mechanically ground and polished to create a flat surface with high optical quality for fluorescence measurement. Some fine scratches were still observed on the surface of the glass as evident in Fig.3a. However these did not have a significant impact on thermometry measurements.

\section{Calibration techniques}
To monitor temperature using EYT glass, calibration measurements on the fluorescence output were performed to establish baseline measurements on temperature using a fixed wavelength excitation at 980 nm. The calibration process is performed using two methods, the collection of the ratiometric fluorescence signal and by integrating the total green fluorescence intensity generated by the $^{2}$H$_{11/2}$ and $^{4}$S$_{3/2}$ states. The ratiometric approach provides high sensitivity that is robust to photon loss\cite{schartner2014fibre,brites2011lanthanide,tang2017study}. Conversely the total fluorescence intensity provides a rapid approach to thermometry when the details of the experiment have not altered since calibration.

\begin{figure}[t]
    \includegraphics[width = \textwidth]{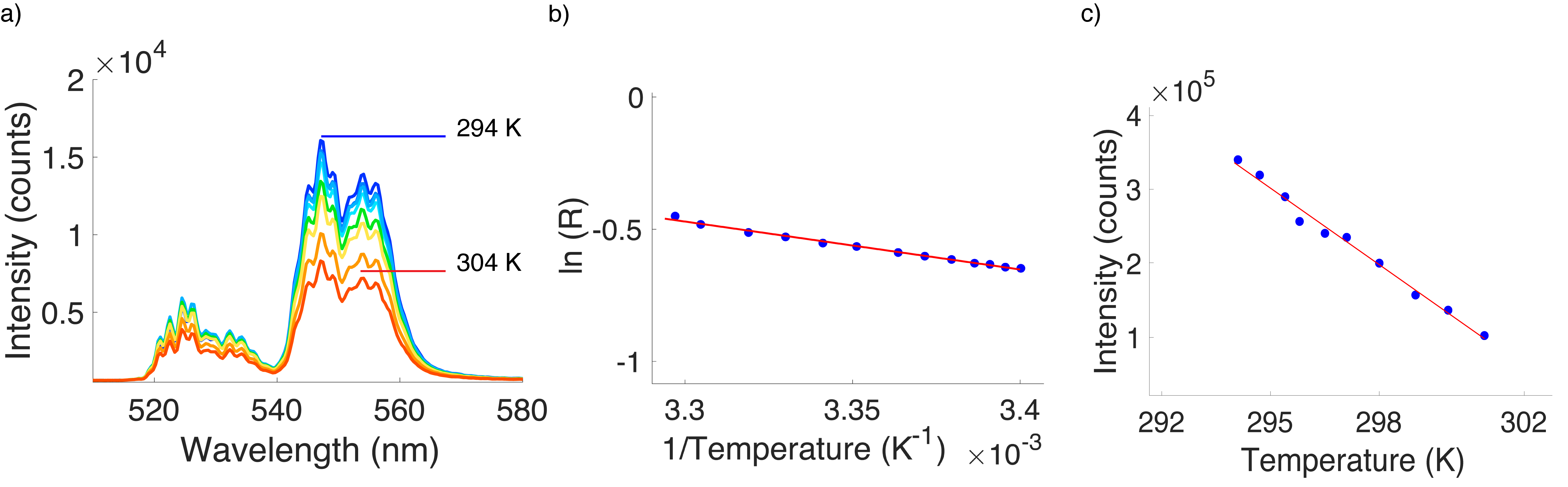}
    \caption{a) Upconversion fluorescence output of the EYT glass used for ratiometric thermometry. Temperature is measured via the variation of emission from the $^{2}$H$_{11/2}$ and $^{4}$S$_{3/2}$ states, where the non-radiative decay rate of the $^{4}$S$_{3/2}$ state increases with temperature yielding a linear trend between emission band ratio and temperature. b) A line of best fit is produced from the measured relationship of temperature and emission band ratio, where a logarithmic fit is applied. c) The total green fluorescence output is calibrated as a function of temperature for fast temperature mapping.}
    \label{fig:2}
\end{figure}

We performed calibration measurements by varying the temperature of the tellurite glass substrate with a Peltier module (Model ZP9104, Jaycar, Australia) and reading the direct temperature of the substrate from an external sensor (Model 408-6109, RS Components, UK). For each specific temperature, spectroscopic information is acquired for 3 seconds at a discrete point with a spot size of $\sim$1.8 $\mu$m that is defined by the excitation wavelength (980 nm) and numerical aperture of the objective lens (Model LCPLN50XIR, 0.65NA, Olympus, Japan), results are shown in Fig. 2a.

The external sensor is attached to the surface of the tellurite glass with thermal compound to ensure an accurate representation of the surface temperature of the glass, which is directly above the Peltier module. For the Peltier module to reach a state of thermal equilibrium a waiting period of 300 seconds per measurement is required, this is only used for calibration purposes as real temperature measurements rely on the glasses response to temperature. The variations in temperature after a 300 second time frame are recorded and used to derive the uncertainty in our calibration measurements which was found to be $\pm$ 0.12 K.

\section{Measurement method}

Measurements on the output signal from the EYT glass are acquired simultaneously by splitting 90 $\%$ towards a spectrometer (SP-2500, Princeton, USA) and 10 $\%$ to an avalanche photodiode detector (APD) (SPCM-AQRH-14, Excelitas, USA).  We are able to measure total fluorescence output as a function of temperature by filtering out the signal from the $^{4}$F$_{9/2}$ photoemission with a 650nm shortpass filter, removing any thermally coupled contributions from this process.

This approach allows for us to interrogate specific regions of interest on the substrate where small scale measurements of temperature is required. To achieve consistent measurements of the upconversion signal, the optimal focal range of the objective lens onto the surface of the substrate was found by identifying the z-position of the scanning stage that yielded the strongest signal (highest counts as measured by the APD). A ratiometric measure of temperature is then calculated from equation 1, with the data shown in Fig. 2b, where we implement a linear fit of the logarithm of the fluorescence intensity ratio with inverse temperature. However, because we used a spectrum analyser to determine the fluorescence spectra, this method has the disadvantage of requiring longer acquisition time periods per point. A faster approach involves monitoring the total green fluorescence output from an APD , which we used to map a 200$\times$200 $\mu$m field of view for continuous temperature measurement. The relationship between green fluorescence intensity and temperature is shown in Fig. 2c, which we use to determine equation 1 showing a linear relationship. From this we are able to measure temperature with a SCM map with a continuous field of view.

Using these calibration methods we are able to determine the temperature of the substrate due to the thermal gradient applied by an object on the surface through equation 1  where $T_{R}$, $T_{F}$, R and F$_{G}$ represent ratiometric temperature, total fluorescence temperature, fluorescence ratio and total green fluorescence intensity.

\begin{eqnarray}
\begin{aligned}
\mathbf{\ln\left( R\right)} & = aT_{R}^{-1} +  b,\quad
\mathbf{F_{G}} & = cT_{F} + d.
\end{aligned}
\end{eqnarray}

The parameters $ a = $-1.8$ \times 10^3~\text{K}^{-1}$ and $b = 5.5$ were determined by using a least squares fit for the ratiometric approach. The parameters for the intensity calibration are $ c = $-3.4$ \times 10^4$ and $d =1 \times 10^7$ counts.

\section{Results and discussion}
To demonstrate the spatially-continuous nature of the measurements, high spatial resolution and timescale, we measured the evaporative cooling of a water droplet by pipetting 0.5 $\mu$L of water on the EYT substrate. The dimensions of the droplet displaced on the substrate is defined by the quantity of water pipetted and also the surface properties of tellurite glass. The hydrophilic nature\cite{yusof2017improved,nurhafizah2017self} of tellurite glass leads to an acute contact angle between the droplet edge and the surface of the EYT thermometer, minimising optical scattering due to the discontinuity at the edge of the droplet.

A scanning confocal map resolving the edge of a water droplet is shown in Fig.3a, given a pixel width of 1 $\mu$m which is defined by the step size of the scanning piezo stage. The area that the droplet covers on substrate corresponds to higher fluorescence when compared to the substrate (room temperature). With this we are able to distinguish between the meniscus of the droplet and small features on the tellurite glass such as small surface cracks (top-left). This intensity map demonstrates the change in temperature of the tellurite glass due to the cooling effect of the droplet, provided by the snapshot in Fig. 3b where we apply an intensity to temperature calibration. In addition to this we measure the ratiometric fluorescence output in order to track the evaporation of the droplet at a single pixel location over a total time period of 180 seconds (Fig. 3c).

\begin{figure}[t]
    \includegraphics[width = \textwidth]{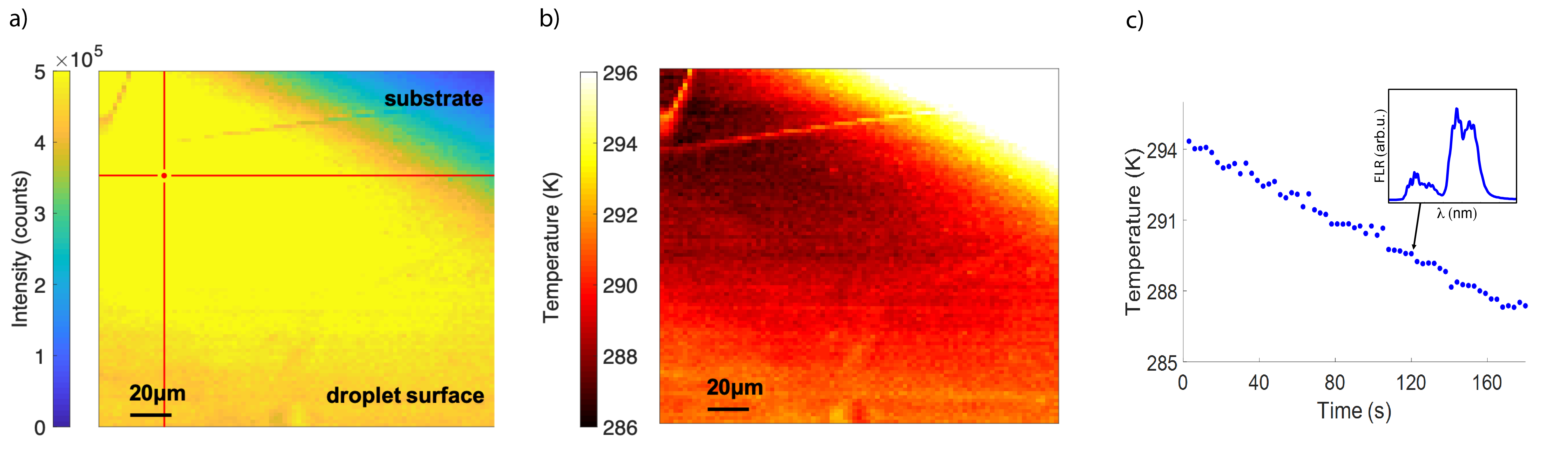}
    \caption{a) A confocal fluorescence intensity map of a water droplet, with an excitation wavelength of 980 nm. The tellurite glass is shown in blue (low fluorescence) and regions displaying higher fluorescence represent the droplet (yellow). b) Inferred temperature map  using fluorescence intensity calibration.  The substrate was at 296 K, corresponding to the ambient temperature, whereas the evaporation of the droplet led to a measurable reduction in temperature under the droplet, with minimum recorded temperature of 286 K. c) By monitoring the fluorescence ratio from the point at which the droplet is placed, a linear decrease in temperature is observed over a time period of 180 seconds. Each data point corresponds to an individual spectrum (inset) after 3 seconds which is then used to find the temperature of the droplet as it begins cooling the tellurite glass.}
    \label{fig:3}
\end{figure}

The measured decrease in temperature is due to evaporative cooling of the droplet, where similar values in literature acquired by using infrared thermography for evaporating water droplets (2 mm in diameter) have been reported to show temperature changes of 6 K\cite{borodulin2016surface}. Evaporation of droplets of this size has also been studied with emission diffusion computational models\cite{borodulin2017determination} that show the total time for droplet evaporation occurs between 20-60 minutes depending on the dimensions of the droplet, humidity, ambient temperature and pressure. We observe a change in size of the droplet as it evaporates, however this process occured at timescales much larger than the single pixel measurement acquired over 180 seconds and is in agreement with the emission diffusion model. By monitoring the change in emission band ratio with 3 second continuous measurements of the tellurite glass, we are able to report a temperature change of 7.04 K $\pm$ 0.12 K. Using this modality, we pave the way for future thermometry applications by demonstrating a proof of concept measurement that is able to acquire real time temperature information over a 200 $\mu$m field of view .

\section{Summary}

In summary we show micrometer resolution temperature sensing with a sensitivity of $\pm$ 0.12 K using EYT glass. The logarithmic trend of fluorescence ratio and linear relationship of total fluorescence intensity are both dependant on temperature and thus can be both used for temperature measurement. This technique provides simultaneous measurements of fluorescence spectra and fast temperature mapping of objects that come into contact with our optical thermometer.  From this we are able to measure and calibrate for the total temperature change of a water droplet embedded on the surface of the tellurite thermometer. Over a time period of 180 seconds we report a temperature change of 7.04 K due to the cooling effect of a water droplet, and build up a temperature map that is able to distinguish thermal variations within the droplet. We aim to use this thermometry modality as a starting point for measuring small scale temperature changes in fundamental processes in biological and chemical systems.

\section*{Acknowledgments}

The authors acknowledge the scientific and technical assistance of the Australian CNBP facilities. This work has been supported by Australian Research Council grants LP150100657, FT160100357, FT180100343 and CE140100003. This work was performed in part at the OptoFab node of the Australian National Fabrication Facility utilizing Commonwealth and SA State Government funding. RAM is supported by a Premier's Research and Industry Fund grant provided by the South Australian Government Department for Industry and Skills.


\bibliography{OE_DS_PAPER2020}






\end{document}